\documentclass[%
 twocolumn,
 amsmath,amssymb,
 aps,
 prb, 
 9pt 
]{revtex4-2}

\usepackage{graphicx}
\usepackage{dcolumn}
\usepackage{bm}
\usepackage[T1]{fontenc}
\usepackage[utf8]{inputenc}
\usepackage{times} 
\usepackage{xcolor}
\usepackage{hyperref}
\usepackage{ulem}
\usepackage{siunitx}

\newcommand{\up}{\uparrow}
\newcommand{\dn}{\downarrow}
\newcommand{\redsout}{\bgroup\markoverwith{\textcolor{red}{\rule[0.5ex]{2pt}{0.4pt}}}\ULon}

\usepackage{titlesec}
\titlespacing*{\section}{0pt}{10pt}{4pt}
\titlespacing*{\subsection}{0pt}{8pt}{3pt}

\usepackage{xr}
\begin{document}
\title{Magnons in metallic altermagnetic $\text{KV}_2\text{Se}_2\text{O}$ }

\author{Daniel Louren\c co R. Santos$^{1}$, Ant\'onio T. Costa$^{2,3}$}

\affiliation{$^1$Centro Federal de Educa\c c\~ao Tecnol\'ogica, 23812-101 Itagua\'i, RJ, Brazil}

\affiliation{$^2$International Iberian Nanotechnology Laboratory, Av. Mestre Jos\'e \ Veiga,\ 4715-330 \ Braga, \ Portugal}

\affiliation{$^3$Centre of Physics of Minho and Porto Universities (CF-UM-UP), Universidade do Minho, Campus de Gualtar, 4710-057 Braga, Portugal}

\begin{abstract}
We investigate the spin excitation in altermagnetic vanadium oxyselenides, focusing on the metallic quasi-two-dimensional compound $\text{KV}_2\text{Se}_2\text{O}$. 
Using a fermionic Hamiltonian derived from ab initio calculations, we compute the transverse spin susceptibilities within the random phase approximation and extract the magnon dispersion relation. We show that the altermagnetic symmetry leads to characteristic degeneracies and directional splittings in both the electronic bands and the magnon spectra along high-symmetry paths of the Brillouin zone. The metallic $\text{KV}_2\text{Se}_2\text{O}$ exhibits finite magnon linewidths arising from the coupling to the particle–hole continuum. Furthermore, by fixing the magnitude of the wave vector and varying its in-plane direction, we uncover a pronounced angular dependence of the magnon energies and lifetimes, with complementary damping behavior between the two magnon branches. Upon including spin–orbit coupling, the system exhibits an out-of-plane easy $c$-axis anisotropy, which opens a magnon gap of approximately 6~meV at the $\Gamma$-point, possibly rendering the magnetic order stable at room temperature. Our results demonstrate that altermagnetism controls not only the dispersion but also the anisotropic decay of spin excitations, highlighting altermagnetic metals as promising platforms for directionally selective magnon transport.\\
\end{abstract}

\maketitle

\section{Introduction}

Collective spin excitations, or magnons, constitute a fundamental aspect of magnetic materials and play a central role in contemporary spintronics and magnonics. As bosonic quasiparticles, magnons enable the transport of spin and angular momentum without charge motion, offering promising routes toward low-dissipation information processing \cite{Chumak2015,Kajiwara2010,Cornelissen2015}. The dispersion, symmetry, and anisotropy of magnon spectra are intimately linked to the underlying magnetic order and crystalline symmetries, making them a powerful probe of magnetic ground states and a key ingredient for functional spin-based devices. Magnon band degeneracies and anisotropic splittings originate from the combined 
spin and space group symmetries in collinear magnets \cite{Liu2025SymmetryMagnons}, while recent studies in 
altermagnetic systems have highlighted chiral magnon modes and direction-dependent magnon behavior tied to 
crystal symmetry and exchange interactions \cite{Kravchuk2025ChiralMagnons, xg1x-sj4c}. Advances in understanding spin 
wave reconstruction and manipulation further emphasize the role of magnon dispersion in device-relevant magnetic dynamics \cite{Wu2025SpinWave,chumak2022advances, Costa_2020}, and comprehensive reviews have underscored these aspects in low-dimensional magnetic materials \cite{Wang2025ReviewMagnons}.

Antiferromagnetic materials have attracted significant attention in spintronics due to their vanishing net magnetization, which suppresses stray fields and enables ultrafast spin dynamics~\cite{Jungwirth2016,Baltz2018,Gomonay2014}. 
However, conventional collinear antiferromagnets (AFMs) are typically constrained by band degeneracies in both electronic and magnonic spectra that limiti their tunability and functional response~\cite{Smejkal2020,Smejkal2022,Smejkal2023PRX}. 
These restrictions have motivated the search for alternative magnetic phases that preserve zero net magnetization while allowing for richer momentum-dependent spin phenomena and controllable spin splitting.

Recently, altermagnetism has emerged as a distinct class of magnetic order that fundamentally extends the conventional dichotomy between ferromagnets and antiferromagnets. Altermagnets exhibit collinear magnetic order with zero macroscopic magnetization, yet their magnetic space-group symmetries allow for a momentum-dependent spin splitting of electronic bands even in the absence of spin–orbit coupling \cite{Smejkal2022,Smejkal2020,Smejkal2023PRX,Feng2022}. 
This unconventional symmetry breaking generates electronic structures that resemble those of ferromagnets, while preserving key advantages of antiferromagnets such as the absence of stray fields and ultrafast spin dynamics, placing altermagnets at the forefront of current research in condensed matter physics and spintronics \cite{Jungwirth2016,Baltz2018}.

Although the electronic structure and transport properties of altermagnets have been extensively 
investigated, their collective spin excitations remain comparatively less explored. 
Most existing studies of spin-wave spectra in altermagnetic systems rely on effective spin 
models, which capture the essential symmetry constraints but often neglect the fully itinerant 
nature of the electronic states~\cite{garcia2025magnon,wu2025magnon,khatua2025magnon}. Recent first-principles investigations have begun to address 
this gap, identifying symmetry-driven features such as chiral magnons and spin–orbit-induced 
effects in two-dimensional altermagnets~\cite{Sodequist2024APL}. Model calculations for
itinerant altermagnets have predicted pronounced directional anisotropies in magnon Landau 
damping arising from coupling to the Stoner continuum~\cite{Costa2025SciPost}, an important 
consequence of the interplay between altermagnetic symmetry and itineracy. 

While collinear AFMs maintain a strict chiral degeneracy between opposite spin magnon modes, the distinct local chemical environments of non-magnetic atoms in altermagnetic materials break the spatial-temporal inversion symmetries that protect degeneracy in AFMs, inducing a strong magnon splitting even in the absence of spin-orbit coupling (SOC). This lifting of degeneracy in momentum space gives rise to unique transport properties, distinguishing altermagnets from both AFMs and systems governed by relativistic mechanisms.\cite{rezende2019introduction,wu2025magnon} Nevertheless, a comprehensive and unified understanding of magnons in altermagnets is 
still lacking. In particular, the fundamental role of electron–hole (Stoner) excitations in itinerant or small-gap altermagnets remains under-explored. 
Addressing these issues is crucial not only for establishing a complete theoretical framework of 
altermagnetic order but also for assessing the suitability of altermagnets as active platforms 
for magnonic and spintronic applications.

 Although the hallmark spin splitting in altermagnets is fundamentally non-relativistic in 
 origin, the inclusion of spin-orbit coupling (SOC) is important: in 2D systems the SOC-induced anisotropy gap in the magnon spectrum determines the very existence of long-range order~\cite{MerminWagner}. SOC also mediates the coupling between unconventional spin-space topologies and itinerant electronic states~\cite{gonzalez2026topological}, and introduces a coupling between transverse spin, longitudinal spin and charge excitations~\cite{costa2010spin}. Consequently, 
 investigating how the magnon dispersion incorporates relativistic corrections provides a  theoretical baseline for interpreting high-resolution spectroscopic measurements and 
identifying the subtle interplay between exchange anisotropy and weak spin-orbit effects.

In this work, we focus on the C-type magnetic configuration of the vanadium oxyselenide $\text{KV}_2\text{Se}_2\text{O}$.
In the C-type configuration, $\text{KV}_2\text{Se}_2\text{O}$ is a quasi-two-dimensional metallic altermagnet, whereas its exfoliated $\text{V}_2\text{Se}_2\text{O}$ monolayer exhibits a semiconducting ground state while maintaining its altermagnetic nature ~\cite{ma2021multifunctional,Jiang2025NP}. Momentum-dependent spin splitting in $\text{KV}_2\text{Se}_2\text{O}$ has been observed by spin- and angle-resolved photoemission spectroscopy (SARPES)~\cite{Jiang2025NP}. Subsequent neutron-diffraction measurements, however, identified a predominantly G-type antiferromagnetic bulk ground state, apparently at odds with the surface-sensitive SARPES observations~\cite{sun2025antiferromagnetic}. Two recent developments help reconcile these findings. First, surface-induced symmetry reduction can remove the bulk symmetry that enforces spin degeneracy, thereby producing an altermagnetic electronic structure at the surface of a G-type antiferromagnet~\cite{lange2026emergent}. Second, recent experiments combining spin-polarized scanning tunnelling microscopy with magnetic-field-dependent quasiparticle-interference imaging have identified both C-type and G-type magnetic configurations in $\text{KV}_2\text{Se}_2\text{O}$~\cite{gu2026}, establishing the experimental relevance of local C-type altermagnetic regions. Our calculations directly predict the magnon excitations of C-type domains and, in the weak-interlayer-coupling limit, provide approximate estimates of the dispersion and Landau damping of surface-dominated spin excitations in G-type domains.

Using first-principles-based fermionic models, we compute the magnon spectra of metallic C-type $\text{KV}_2\text{Se}_2\text{O}$ and analyze their energies and lifetimes in terms of anisotropies, degeneracies, and symmetry-imposed constraints. We assess the effects of SOC on their respective magnon spectra and find that, while it opens a sizable anisotropy gap at the $\Gamma$-point, it has little effect on other features of the magnon spectrum, thus preserving the features that characterize it as altermagnetic.

The remainder of this paper is organized as follows. In Sec.II, we describe the computational methodology, including the \textit{ab initio} electronic structure calculations and the construction of the multi-orbital tight-binding Hamiltonians using the PAOFLOW framework. In Sec.~III, we discuss the electronic structure of C-type $\text{KV}_2\text{Se}_2\text{O}$. We then present the transverse spin susceptibility calculated within the random phase approximation (RPA) to examine the magnon dispersion relations and evaluate the effects of spin-orbit coupling (SOC). Furthermore, we analyze the angular dependence of the magnon spectrum in the metallic phase, emphasizing the anisotropic damping and finite lifetime of these spin excitations. Finally, Sec. IV summarizes our main conclusions and outlines possible implications for altermagnetic magnonics.

\section {Methodological framework}

The magnon spectra are obtained using an effective fermionic model whose parameters are extracted from first-principles calculations. In the following, we outline the computational procedure employed to construct the empirical Hamiltonian and to evaluate the spin-wave excitations.

\subsection{Ab initio calculations}

First-principles calculations were performed within density functional theory (DFT) using the Quantum ESPRESSO package \cite{giannozzi2009quantum, giannozzi2017advanced}. Different exchange–correlation treatments were adopted for the bulk and monolayer systems in order to accurately capture their distinct electronic characters and to ensure consistency with available experimental and theoretical results.

The quasi-two-dimensional bulk compound $\text{KV}_2\text{Se}_2\text{O}$, correlation effects were treated at the GGA level without an explicit Hubbard $U$ term. The resulting GGA band structure accurately reproduces previously reported theoretical results and is consistent with spin- and angle-resolved photoemission spectroscopy (SARPES) measurements \cite{Jiang2025NP,xu2025electronic}. This approach allows us to describe both systems within a unified framework while faithfully capturing their experimentally established electronic properties.

\subsection{Ab initio derived tight-binding Hamiltonian}

Based on the spin-polarized ab initio electronic structure, we construct a multi-orbital tight-binding Hamiltonian of the form
\begin{equation}
H = H_0 + H_{\mathrm{SOC}} ,
\end{equation}
where $H_0$ describes the single-particle hopping processes and contains the mean-field exchange fields associated with local magnetic moments. $H_{\mathrm{SOC}}$ introduces relativistic spin-orbit effects. The single-particle term is given by
\begin{equation}
H_0 = \sum_{ij}\sum_{\mu\nu,\sigma} T_{ij}^{\sigma\mu\nu}
c^{\dagger}_{i\mu\sigma} c_{j\nu\sigma},
\end{equation}
where $c^\dagger_{i\mu\sigma}(c_{i\mu\sigma})$ creates (annihilates) an electron with spin $\sigma$ orbital $\mu$ at the site $i$. The indices $\mu$ and $\nu$ run over one $s$, three $p$, and five $d$ orbitals for each vanadium site, as well as the valence $s$ and $p$ orbitals for both selenium and oxygen sites. The hopping parameters $T_{ij}^{\sigma\mu\nu}$ are obtained by projecting the DFT Hamiltonian onto a pseudo-atomic orbital (PAO) basis using the PAO projection technique~\cite{nardelli2018paoflow}. The diagonal terms $T_{ii}^{\sigma\mu\nu}$ contain, among other contributions, the exchange field associated with local magnetic moments $m_i\equiv\sum_{\mu}\langle c^\dagger_{i\mu\up}c_{i\mu\up} - c^\dagger_{i\mu\dn}c_{i\mu\dn}\rangle$. To connect this spin-polarized Hamiltonian
with our approach to calculate the magnon spectra~\cite{costa2010spin,Costa_2020}, we need to associate the 
spin-polarized DFT-derived PAO Hamiltonian to the mean-field approximation
derived from the multi-orbital Hubbard term
\begin{equation}
H_U = \sum_{i \in \text{V}} \sum_{\mu\nu\mu'\nu'} \sum_{\sigma\sigma'}
U^{\mu\nu\mu'\nu'}_i
c^{\dagger}_{i\mu\sigma}
c^{\dagger}_{i\nu\sigma'}
c_{i\mu'\sigma'}
c_{i\nu'\sigma},
\end{equation}
where the site index $i$ runs solely over the vanadium positions, $U_i^{\mu\nu\mu'\nu'}$ denotes 
the elements of the on-site screened Coulomb interaction, acting only on the 3d orbitals of 
vanadium. In our approach, we choose the simplest possible parametrization for the matrix 
elements of the interaction, $U^{\mu\nu\mu'\nu'}=I\delta_{\mu\nu'}\delta_{\nu\mu'}$. Importantly, $H_U$ is not added to the DFT-derived Hamiltonian $H_0$, nor is it subjected to an additional mean-field decoupling. The static exchange field associated with the magnetic reference state is already contained in the spin-dependent matrix elements of $H_0$.
The connection is established by writing
\begin{equation}
    T_{ii}^{\sigma\mu\mu} = \varepsilon_i^\mu  + \tau_\sigma\frac{Im_i}{2}
\end{equation}
where $i\in \text{V}$ and $\tau_\sigma=-1,1$ for $\sigma=\up,\dn$.
The effective interaction strength $I$ (which should not be confused with the Hubbard $U$ employed 
in the GGA+$U$ electronic structure calculations) is then determined by enforcing the Goldstone condition in the absence of SOC,
\begin{equation}
    \mathrm{det}\left[\mathbf{1}+I\Bar{\boldsymbol{\chi}}^{+-}(\vec{q}=\vec{0},\omega=0)\right] = 0
\end{equation}
where $\Bar{\boldsymbol{\chi}}^{+-}$ is the mean-field transverse spin susceptibility matrix obtained from $H_0$ projected on the vanadium 3D subspace~\cite{lounis2010dynamical,khajetoorians2011itinerant,lounis2011theory}. 
The resulting value is 
$I = 0.91$~eV for Vanadium atoms in $\text{KV}_2\text{Se}_2\text{O}$. The value obtained in the metallic system is consistent with effective screening \cite{csacsiouglu2012strength}. The same values are retained when $H_{\mathrm{SOC}}$ is subsequently included, allowing the SOC-induced anisotropy gap to emerge without further adjustment.

Finally, the spin--orbit coupling (SOC) is included as
\begin{equation}
H_{\mathrm{SOC}} =
 \sum_i \sum_{\alpha,\beta} \sum_{\sigma,\sigma'}
\lambda_i\ c^{\dagger}_{i\alpha\sigma}
\left( \mathbf{L}_{\alpha\beta} \cdot \mathbf{S}_{\sigma\sigma'} \right)
c_{i\beta\sigma'},
\end{equation}

where $\lambda_i$ denotes the atomic SOC strength at site $i$, $\mathbf{L}_{\alpha\beta}$ are the matrix elements of the orbital angular momentum operator in the orbital basis, and $\mathbf{S}_{\sigma\sigma'}$ are the spin-$1/2$ Pauli matrices. The atomic spin-orbit coupling parameters for the constituent elements are set to $\lambda_{\text{V}} = \text{22}$ meV for vanadium and $\lambda_{\text{Se}} = \text{206}$ meV for selenium, with both values taken from Ref.~\cite{montalti2006handbook}. Crucially, while the intrinsic SOC coefficient for the vanadium $3d$ shell ($\lambda_{\text{V}}$) is minimal, the relativistic effects in this compound are predominantly driven by the heavy selenium $4p$ ligands ($\lambda_{\text{Se}}$). Through the $d-p$ hybridization embedded in $H_0$, this ligand-assisted SOC is effectively transferred to the itinerant magnetic channels. One virtue of our model Hamiltonian is that it allows us to separate relativistic and non-relativistic effects by dialing down the SOC strength $\lambda_X$, while still faithfully reproducing the relativistic DFT results when the full values of $\lambda_X$ is used.

Our \textit{ab initio} and PAOFLOW calculations show that the characteristic altermagnetic band splittings and symmetry-induced degeneracies are already fully captured in the absence of SOC ($\lambda=0$). This is consistent with the nonrelativistic origin of altermagnetism, which arises from the combination of crystal symmetry and collinear antiferromagnetic order rather than from relativistic effects. Moreover, for vanadium-based $3d$ systems, SOC is expected to be weak and to introduce only higher-order corrections, such as small anisotropy gaps, without qualitatively modifying the magnon dispersion or damping. Neglecting SOC therefore allows us to isolate the dominant altermagnetic mechanism governing the spin excitations while keeping the analysis transparent.

This fermionic model Hamiltonian provides a framework to describe the interplay between electron itinerancy, electronic correlations, and spin--orbit coupling, which are essential ingredients for the emergence of altermagnetic order and the resulting magnon excitations, all of that in the context of a realistic electronic band structure.

\subsection{Spin susceptibility and magnon spectrum}

The collective spin excitations are described in terms of the transverse spin susceptibility, which governs the nonlocal spin response of the system to an external magnetic perturbation. Within linear response theory, the spin susceptibility is defined as the retarded correlation function between spin density operators, encoding both the spatial and temporal propagation of spin fluctuations. 
\begin{equation}
\chi^{\perp}_{ab}(\mathbf{r},\mathbf{r}',t)
=
-i\,\theta(t)\,
\left\langle
\left[
\hat{S}^{\eta}_{a}(\mathbf{r},t),
\hat{S}^{\eta'}_{b}(\mathbf{r}',0)
\right]
\right\rangle ,
\end{equation}
Here, $\perp=\eta\eta'=+-$ or $-+$ label the transverse spin components defined by the ladder operators $\hat{S}^{\pm} = \hat{S}^x \pm \hat{S}^y$, while the indices $a$ and $b$ denote sublattice degrees of freedom within the unit cell at position ${\mathbf{r}}$ and time $t$ relative to the $z$-axis magnetic quantization~\cite{costa2006spin,costa2010spin}. The poles of the transverse spin susceptibility seen as a function of $\omega$ for fixed wave vectors determine the magnon spectrum. In the present work, the interacting susceptibility matrix is evaluated within the random phase approximation (RPA), which captures the coupling between collective spin modes and itinerant electron–hole excitations, 
\begin{equation}
\boldsymbol{\chi}^{\perp}(\mathbf{q},\omega) =
\left[
\mathbf{1} + I  \boldsymbol{\bar{\chi}}^{\perp}(\mathbf{q},\omega)
\right]^{-1}
\boldsymbol{\bar{\chi}}^{\perp}(\mathbf{q},\omega),
\end{equation}
where $\boldsymbol{\bar{\chi}}$ is the mean-field transverse susceptibility matrix projected on the vanadium 3d subspace and $I$  is the effective on-site Coulomb interaction strength discussed in the previous section~\cite{costa2006spin,costa2010spin}.
The magnon dispersion relations are finally obtained by locating the poles of the transverse RPA susceptibility $\boldsymbol{\chi}^{+-}(\mathbf{q},\omega)$ and $\boldsymbol{\chi}^{-+}(\mathbf{q},\omega)$ at finite wave vectors, which correspond to the collective spin-wave excitations of the altermagnetic state. In both conventional antiferromagnets and altermagnets, the collective nature of the magnon modes leads to characteristic peaks in the spectral densities
\begin{equation}
    \Lambda^{\perp}_i(\vec{q},\omega)\equiv-\frac{1}{\pi}\mathrm{Im}\chi^{\perp}_{ii}(\vec{q},\omega),
\label{eq:Lambda}
\end{equation}
with $\perp=+-$ or $\perp=-+$, at the resonance energies of both modes on both sites. The spectral weight distribution of a given mode indicates the precession amplitudes associated with each sublattice. On the other hand, whenever the modes are non-degenerate, $\Lambda^{+-}$ and $\Lambda^{-+}$ will have a single peak each, at different energies. This means that the modes can be labeled by their $S^z$ value, $+\hbar$ for $\Lambda^{-+}$ and $-\hbar$ for $\Lambda^{+-}$. For simplicity, we will refer to these modes as having $S^z=\pm 1$. It is common to see  in the literature these modes being described as having different and well-defined chiralities, which is equivalent to our description in terms of ther $S^z$ quantum number \cite{Costa2025SciPost}.

\section{Results}

\begin{figure}[!htbp]
\includegraphics[width=0.27\textwidth]{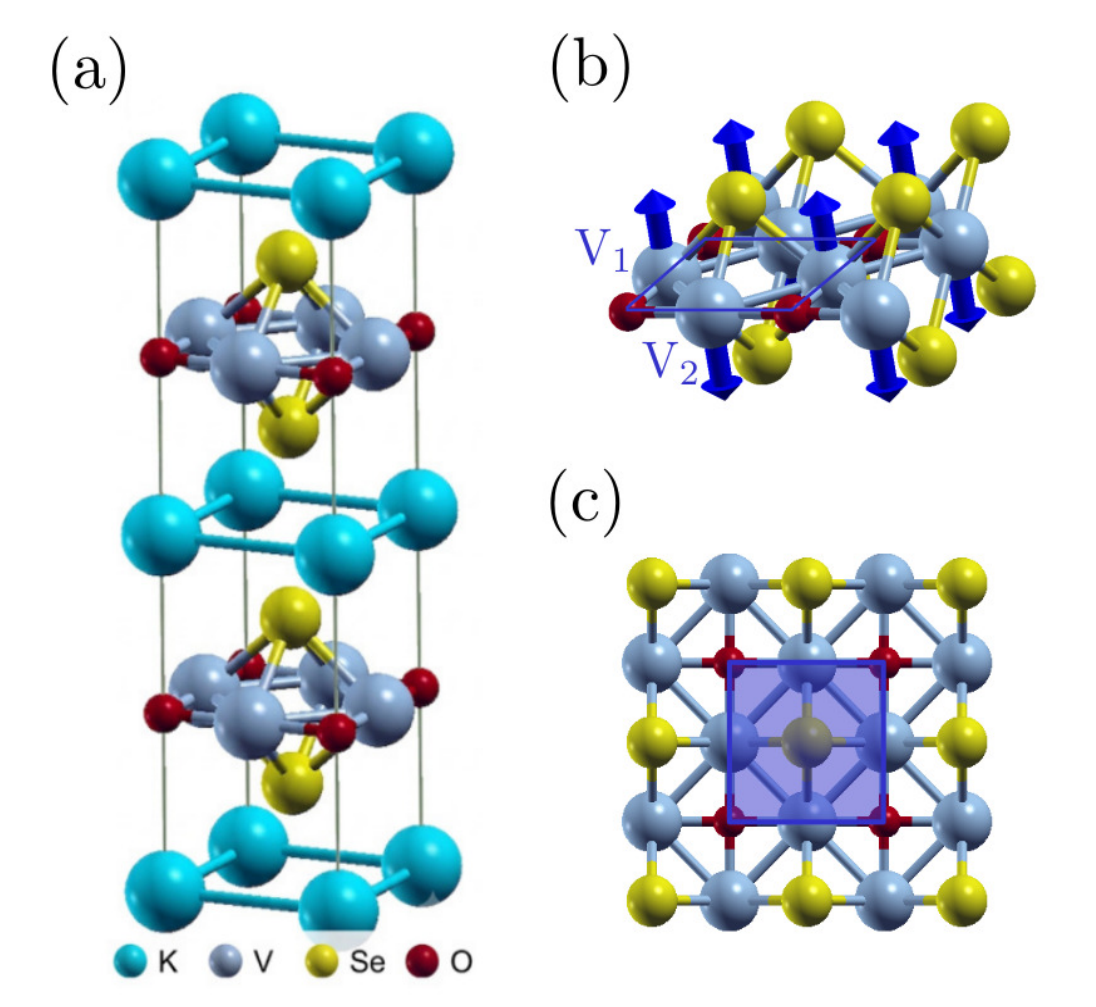}
\hspace{0.1 cm}
    \includegraphics[width=0.3\textwidth]{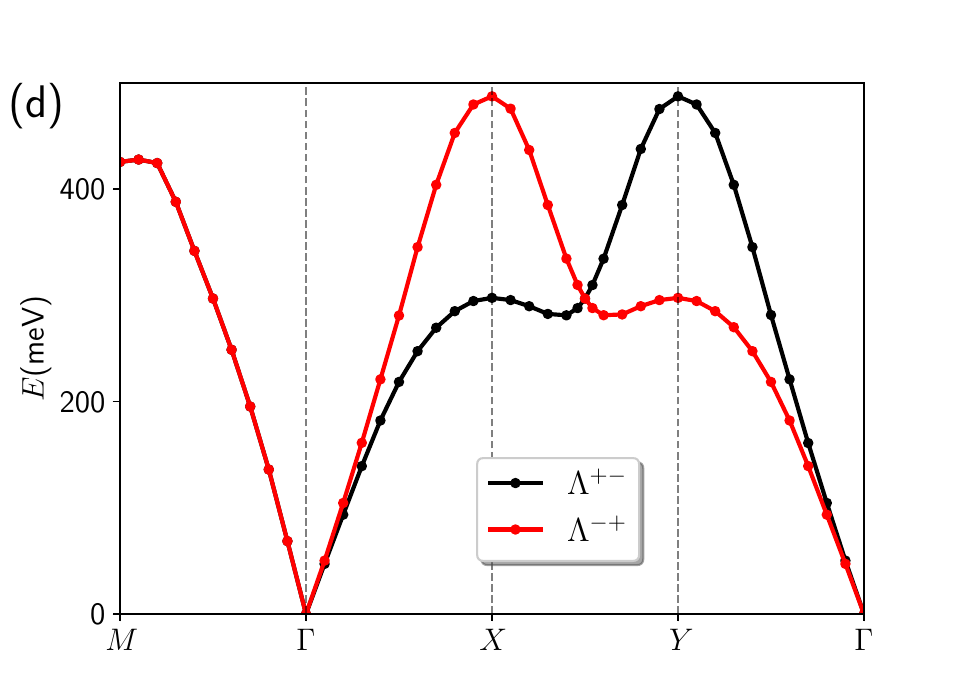}
    \hspace{0.1 cm}
    \includegraphics[width=0.3\textwidth]{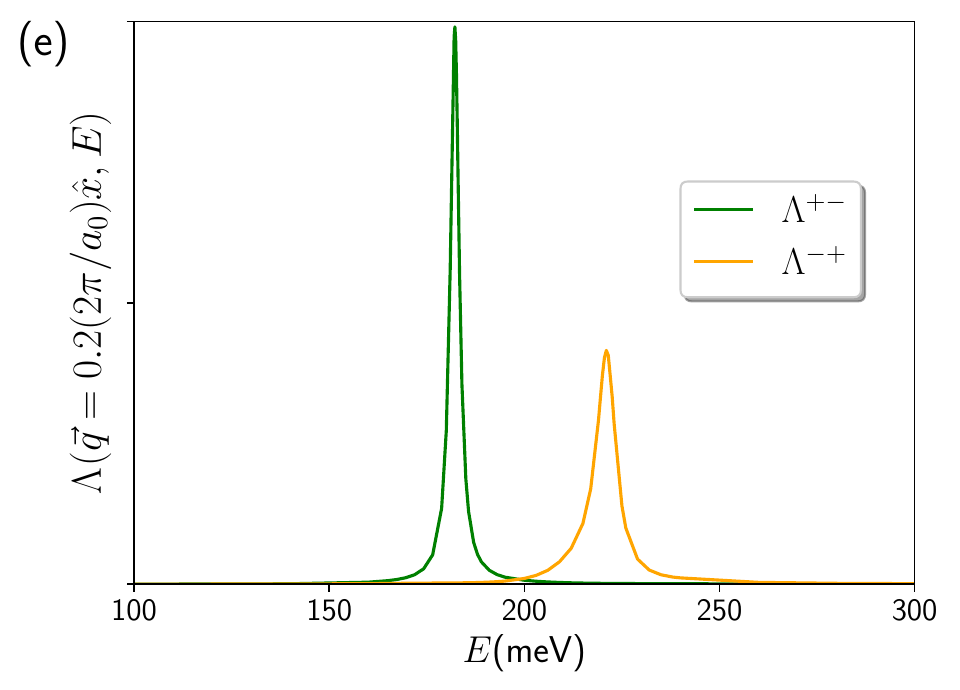}
  	\caption{\label{fig:fig1} (a) View of the bulk quasi-two-dimensional $\text{KV}_2\text{Se}_2\text{O}$ structure, highlighting the layered arrangement and the vanadium magnetic sublattices. (b) Atomic structure of the monolayer $\text{V}_2\text{Se}_2\text{O}$ obtained by exfoliating. The two atoms non-equivalents are labeled $\text{V}_1$ and $\text{V}_2$, which host antiparallel magnetic moments in the altermagnetic state. (c) Top view, the square lattice symmetry is emphasized. (d) Magnon dispersion of the metallic bulk $\text{KV}_2\text{Se}_2\text{O}$ along the high-symmetry path $M$–$\Gamma$–$X$–$Y$–$\Gamma$. (e) Calculated spectral densities $\Lambda^{+-}$ and $\Lambda^{-+}$, as a function of excitation energy $E$ for fixed wave vector $\textbf{q}=0.2\ (2\pi/a_0)\hat{x}$}
\end{figure}

\textit{}{ Altermagnetic order:} 
We begin by discussing the magnetic ground state and crystal symmetry of $\text{KV}_2\text{Se}_2\text{O}$, which form the basis for the altermagnetic behavior explored in this work. This systems presents a collinear antiferromagnetic order with vanishing net magnetization, characterized by two magnetic vanadium sublattices within the crystallographic unit cell. These sublattices, denoted as V$_1$ and V$_2$, host magnetic moments that are antiparallel to each other, as illustrated in Fig. \ref{fig:fig1}.

Figure \ref{fig:fig1}(a) shows the C-type $\text{KV}_2\text{Se}_2\text{O}$ where the vanadium atoms form a layered square lattice with two magnetic sites per unit cell. For a better visualization of magnetic ground-state of $\text{KV}_2\text{Se}_2\text{O}$, the Fig. \ref{fig:fig1}(b) present the exfoliated monolayer $\text{V}_2\text{Se}_2\text{O}$, where the same magnetic arrangement is preserved. The reduction to two dimensions does not modify the underlying collinear antiferromagnetic order. The crystal structure possesses square lattice symmetry, which would conventionally imply equivalent electronic dispersions along symmetry-related directions in the Brillouin zone.

Before addressing the spin excitation spectra, we first assess the reliability of the electronic structure obtained from the PAOFLOW-derived tight-binding Hamiltonians (Further computational details regarding the electronic structure calculations are provided in the Supplemental Material (SM)~\cite{SM}). For the bulk compound, our calculations reproduce the metallic character and the main dispersive features reported SARPES measurements, in agreement with previous theoretical studies  ~\cite{Jiang2025NP}. In particular, the relative positions, bandwidths, and Fermi-level crossings of the vanadium-derived bands closely match those extracted from SARPES, confirming that the low-energy electronic structure is faithfully captured without invoking an on-site Hubbard correction. The resulting PAOFLOW bands are indistinguishable from the Quantum ESPRESSO eingenvalues, precisely capturing the gap opening and overall dispersion.
The characteristic altermagnetic spin splitting is clearly observed. In particular, the spin character of the bands is inverted between the $\Gamma$-X and $\Gamma$-Y directions: a band that is predominantly spin-down along $\Gamma$-X becomes predominantly spin-up along $\Gamma$-Y, and vice versa (see SM~\cite{SM}) . 

Although the bands remain degenerate in energy due to the combined crystal and magnetic symmetries, their spin-resolved components exhibit opposite polarization along orthogonal momentum directions. This momentum-dependent spin texture constitutes a hallmark of altermagnetism and directly reflects the presence of two antiferromagnetically coupled magnetic sublattices within a square-symmetric crystal environment.

\textit{Magnon spectrum: }
Figure \ref{fig:fig1}(d) presents the magnon dispersion relations of $\text{KV}_2\text{Se}_2\text{O}$ along the high-symmetry path $M$–$\Gamma$–$X$–$Y$–$\Gamma$. As discussed above, The magnon dispersion is obtained by mapping the $\omega$ values at which the sublattice-projected spectral densities $\Lambda_i^\perp(\omega,\vec{q})$, Eq.~\ref{eq:Lambda}, peak for each $\vec{q}$; $\perp$ stands for either $+-$ or $-+$, associated respectively with $S^z=-1$ or $S^z=+1$ magnons. The altermagnetic order gives rise to characteristic splittings and degeneracies of the magnon branches along symmetry-related directions. Notably, the magnitude of the altermagnetic splitting is large (reaching roughly 50\% of the largest magnon energies), reflecting the strong spin-dependent electronic band asymmetry and the large effective exchange in $\text{KV}_2\text{Se}_2\text{O}$.

To quantify the finite lifetime and spectral broadening of the collective spin modes, Fig.~\ref{fig:fig1}(e) displays the magnon spectral densities evaluated at a fixed wave vector. Pronounced spectral linewidths are observed, reflecting the substantial damping induced by the coupling of magnons to single-particle spin-flip excitations within the Stoner continuum — a mechanism directly analogous to the Landau damping of plasmons.

\begin{figure}[!htbp]
    \includegraphics[width=0.35\textwidth]{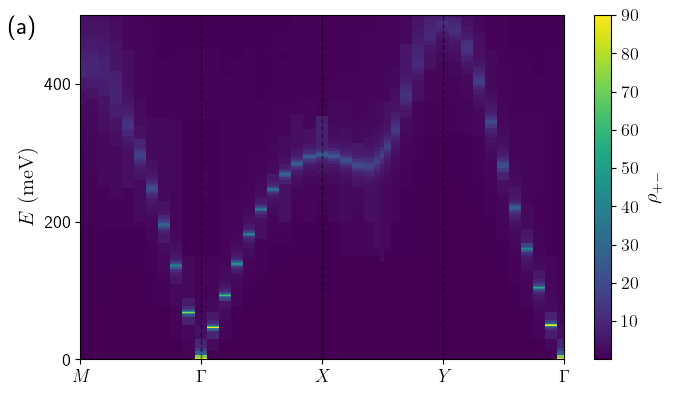}
    \vspace{0.1 cm}
    \includegraphics[width=0.35\textwidth]{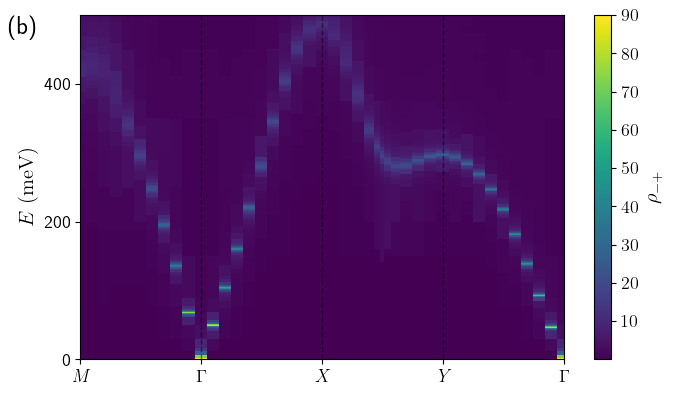}
    
  	\caption{\label{fig:fig2} Color maps of the magnon spectral density  $\Lambda^\perp(\mathbf{q},\omega)$, for the first (a) and second (b) magnon modes in the bulk compound, highlighting the finite linewidth and anisotropic damping induced by the coupling to particle–hole excitations.}
\end{figure}

In altermagnets, Landau damping of magnons is affected by the altermagnetic symmetry, leading to a peculiar damping pattern across the Brillouin zone. This is illustrated in Figs. \ref{fig:fig2}(a) and \ref{fig:fig2}(b), where color maps of the spectral densities, given by Eq.~\ref{eq:Lambda}, are shown for the first and second magnon modes of the bulk system. A finite linewidth is clearly seen throughout the Brillouin zone as a consequence of the coupling between magnons and  Stoner excitations. As expected, the linewidth is strongly suppressed in the vicinity of the $\Gamma$ point, where long-wavelength spin excitations are protected by symmetry and phase-space restrictions, leading to longer-lived magnons. 
Away from $\Gamma$, the increasing overlap with the Stoner continuum results in enhanced damping, whose momentum dependence reflects the underlying altermagnetic electronic structure.

\begin{figure}[!htbp]
    \includegraphics[width=0.35\textwidth]{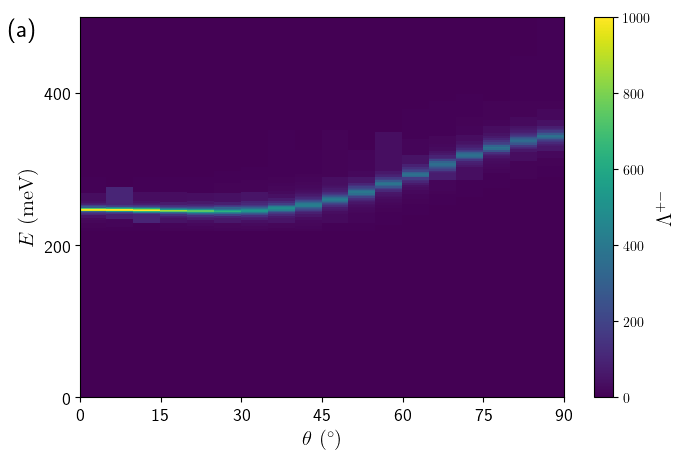}
    \vspace{0.1 cm}
    \includegraphics[width=0.35\textwidth]{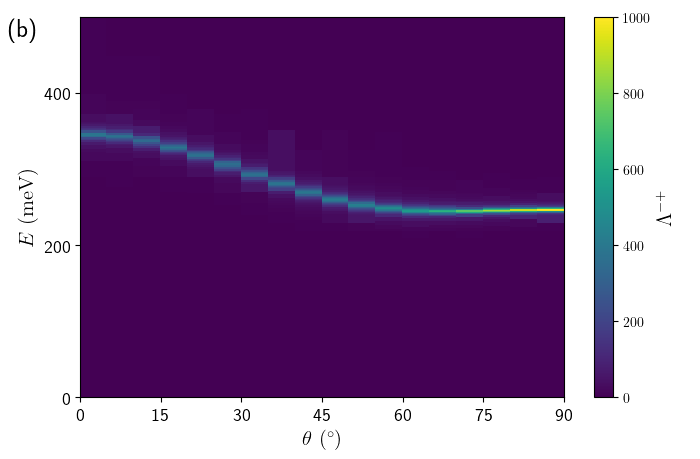}

  	\caption{\label{fig:fig3} Color maps of the magnon spectral density  $\Lambda^\perp(\mathbf{q},\omega)$, for the first (a) and second (b) magnon modes in the bulk compound, highlighting the finite linewidth and anisotropic damping induced by the coupling to particle–hole excitations.}
\end{figure}

\textit{Angular dependence of altermagnetic magnon excitations:}
Model calculations~\cite{Costa2025SciPost} predicted that the lifetimes of magnons in metallic altermagnets should vary strongly with propagation direction. To verify this claim within the context of a realistic, DFT-based calculation, we analyze the magnon spectral density for a fixed wave vector length $|\mathbf{q}| = 0.3(2\pi/a_0)$ as a function of $\theta\equiv\arccos(\hat{q}\cdot\hat{x})$. In Fig.~\ref{fig:fig3} we show density plots of $\Lambda^{+-}_1$ and $\Lambda^{-+}_2$ as functions of $\theta$ and excitation energy. As expected from symmetry considerations, the two modes exhibit a mirror-like behavior upon angular rotation, reflecting the underlying square lattice symmetry and the altermagnetic spin structure.

For the $S^z=-1$ mode, associated with $\Lambda^{+-}_1$, we observe in Fig. \ref{fig:fig3}(a) that at small angles the excitation energy remains nearly constant and the linewidth is weakly dependent on $\theta$, indicating long-lived magnons. As the angle increases, however, the mode shifts to higher energies and acquires a substantially enhanced linewidth, signaling stronger damping due to increased coupling to Stoner excitations. In contrast, the $S^z=+1$ mode, associated with $\Lambda^{-+}_2$, the Fig. \ref{fig:fig3}(b) displays complementary behavior: its energy and linewidth are nearly angle-independent for angles close to $\theta = \ang{90}$, while pronounced broadening and energy renormalization occur at smaller angles.

This complementary angular dependence demonstrates that altermagnetic symmetry not only controls the dispersion of spin excitations but also leads to a highly anisotropic magnon lifetime. Such behavior goes beyond the predictions of linear spin wave theory based on localized Heisenberg models, for which magnons are intrinsically undamped, and highlights the essential role of itinerant electrons captured within the RPA framework.

\textit{Relativistic effects:}
The overall effect of SOC on the electronic structure is relatively weak. The most noticeable modification is the lifting of band crossing points near the $\Gamma$ (see Supplemental Material \cite{SM}). In the vicinity of the Fermi level, which dominates the behavior of the Green's functions and the low-energy spin excitations, the band structure remains largely unaffected. This indicates that SOC does not significantly alter the low-energy electronic states responsible for the magnetic response.
The magnetic anisotropy energy (MAE) calculated via DFT using the force theorem confirms that the ground state possesses an out-of-plane magnetic polarization, with the magnetic moments aligning along the $z$-axis (perpendicular to the planes). The resulting anisotropy energy is small, $\mathrm{MAE} \approx 0.1\text{ meV}$ \cite{Jiang2025NP,li2025magnetic,yan2026magnetic}.

\begin{figure}[!htbp]
   	\includegraphics[width=0.35\textwidth]{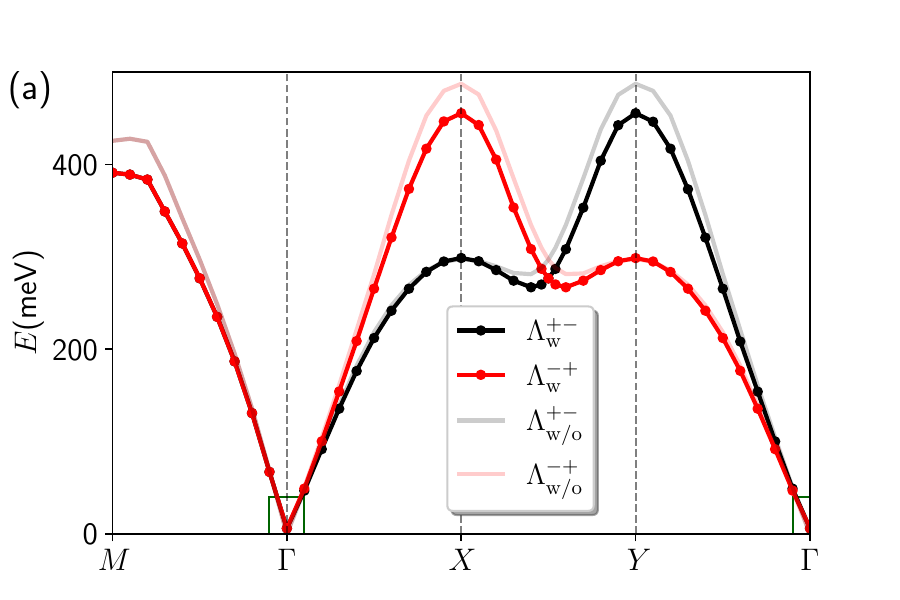}
   	\includegraphics[width=0.35\textwidth]{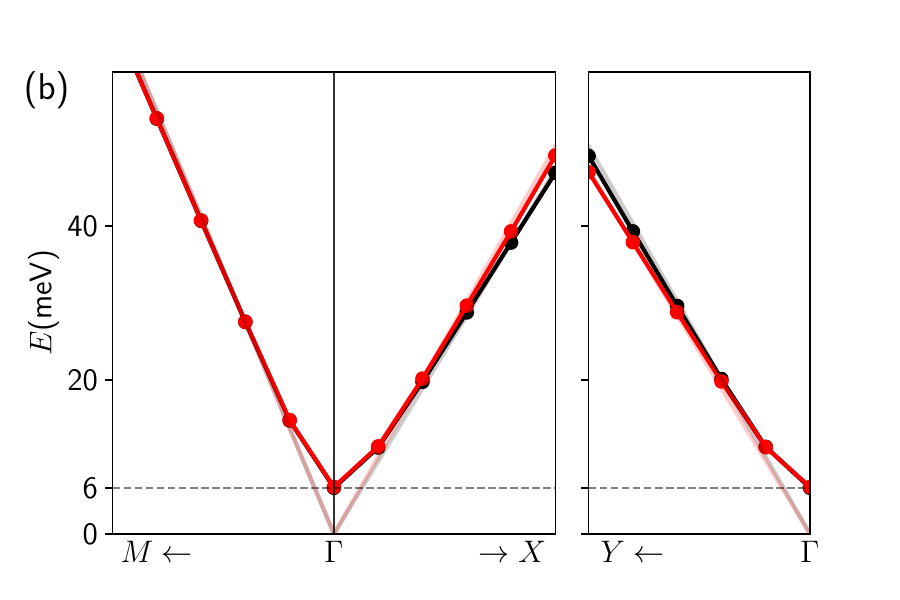}
   	\includegraphics[width=0.35\textwidth]{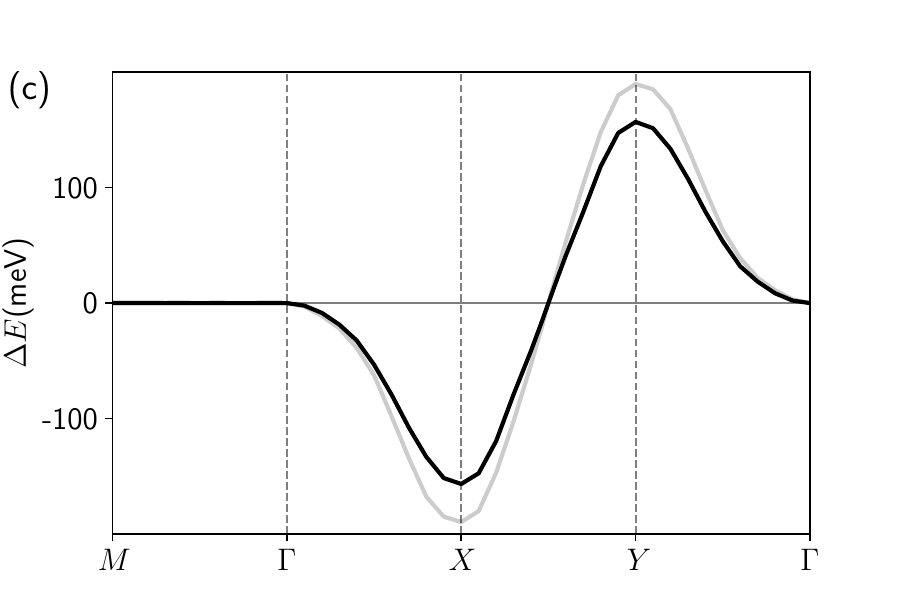}
    \caption{\label{fig:fig4} (a) Transverse magnon dispersion relation. The two magnon branches with SOC are shown as solid lines (black for the first mode and red for the second mode), while the corresponding dispersions without SOC are represented as shaded curves. (b) A magnified view of the low-energy region in the vicinity of the $\Gamma$ point [highlighted by green rectangles in (a)]. (c) The calculated magnon energy splitting as a function of momentum. The grey curve represents the pure non-relativistic splitting, while the black curve displays the case with SOC. }
\end{figure}

To assess the impact of relativistic effects on the collective spin dynamics, we calculate the magnon dispersion relation for the altermagnetic phase of $\text{KV}_2\text{Se}_2\text{O}$ both with and without the inclusion of spin-orbit coupling (SOC). The two main effects are a anisotropy gap at the $\Gamma$-point ($\Delta_\Gamma\approx 6$~meV) and a slight reduction of the altermagnetic splitting between the two magnon branches, shown in Fig.~\ref{fig:fig4}(a). The large anisotropy gap is not surprising: the magnon gap for Néel-type spin arrangements scales not only with the anisotropy, but also with the exchange coupling~\cite{rezende2019introduction}. This gap is, however, essential to ensure the stability of long-range order at finite temperatures. A critical temperature can be roughly 
estimated from the gapped magnon dispersion relation by fitting it to a spin model and using an extension of the approach of Ref.~\cite{Callen1963} to a two-sublattice ferrimagnet. The details of the approach are given in the Supplementary Material \cite{SM}. The estimated critical temperature for 
$\text{KV}_2\text{Se}_2\text{O}$ is $T_N\sim 450$~K. 

The altermagnetic splitting $\Delta E(\vec{q}) = E^{+-}(\vec{q}) - E^{-+}(\vec{q})$ is plotted in Fig.~\ref{fig:fig4}(c) along the high-symmetry paths in the Brillouin zone, with and without SOC. The effect of SOC is to reduce the splitting, with a maximum reduction of $\sim 30\%$. Thus, while SOC makes altermagnetism in $\text{KV}_2\text{Se}_2\text{O}$ potentially stable at room temperature, its deleterious effect on the altermagnetic magnon splitting is modest, preserving the hallmark feature of altermagnetic excitations.

\section{Conclusions}

In summary, we have presented a comprehensive ab initio study of the collective spin dynamics in the metallic quasi-two-dimensional altermagnet $\text{KV}_2\text{Se}_2\text{O}$. By combining PAOFLOW-derived multi-orbital Hamiltonians with the random phase approximation, we established how the characteristic direction-dependent spin splittings of the altermagnetic state directly map onto the magnon dispersions. In this metallic system, the interaction between collective spin waves and the continuum of itinerant particle–hole excitations imparts a finite linewidth to the magnons. Crucially, our realistic first-principles treatment demonstrates that altermagnetism induces a pronounced angular anisotropy not only in the magnon frequencies but also in their Landau damping.

The incorporation of relativistic effects lifts non-relativistic degeneracies, opening a spin-wave gap of $\approx 6.0\text{ meV}$ ($\sim 1.4\text{ THz}$) at the $\Gamma$ point and driving a finite magnon spin splitting of $\approx 0.1\text{ meV}$ ($\approx 24\text{ GHz}$) along the structurally protected $\Gamma$--$M$ nodal line. Importantly, these fundamental energy scales position the collective excitations of $\text{KV}_2\text{Se}_2\text{O}$ squarely within the operational GHz-to-THz frequency regime
demanded by next-generation high-speed spintronic applications. These findings provide a quantitative microscopic foundation for understanding magnonic dissipation and transport in metallic altermagnets. Beyond establishing $\text{KV}_2\text{Se}_2\text{O}$ as a prototypical platform for high-frequency spin dynamics, our work indicates that exploiting the angular control of magnon lifetimes opens promising avenues for dissipation engineering and spin-wave manipulation in altermagnetic spintronic devices.

\section*{Acknowledgments}
 The authors acknowledge the National Laboratory for Scientific Computing (LNCC/MCTI, Brazil) for providing HPC resources of the SDumont supercomputer, which have contributed to the research results reported within this paper (URL: \url{http://sdumont.lncc.br}). This work also used resources of the Centro Nacional de Processamento de Alto Desempenho em S\~{a}o Paulo (CENAPAD-SP). ATC acknowledges the use of Marenostrum5 at Barcelona Supercomputing Centre, through FCT grants 2025.00090.CPCA and 2025.13718.CPCA.

\bibliographystyle{apsrev4-2}
\bibliography{bib}

\appendix

\section{Ab initio details}

The interaction between valence electrons and ionic cores was described using projected augmented wave (PAW) pseudopotentials, and a plane-wave energy cutoff of 80 Ry was employed. Brillouin-zone integrations were performed using Monkhorst–Pack $k$-point meshes of $21 \times 21 \times Nk_z$ for $\text{KV}_2\text{Se}_2\text{O}$. We explicitly verified that the electronic bands projected onto the in-plane Brillouin zone are insensitive to the choice of $Nk_z$. When resolved along the out-of-plane direction, the bands display negligible dispersion and are essentially flat, confirming the quasi-two-dimensional electronic character of $\text{KV}_2\text{Se}_2\text{O}$. {This behavior is consistent with previous first-principles studies of the bulk compound \cite{Jiang2025NP,xu2025electronic}. The converged electronic structures obtained in this manner were used as input for the construction of the effective tight-binding Hamiltonians. This quasi-two-dimensional character provides a natural framework for comparing bulk and monolayer altermagnetic spin excitations on equal footing.

 We first assess the reliability of the electronic structure obtained from the PAOFLOW-derived tight-binding Hamiltonians. Figure~\ref{fig:fig_suppl1} displays the spin-resolved electronic band structure of bulk $\text{KV}_2\text{Se}_2\text{O}$ along the $\Gamma$–$X$–$M$–$Y$–$\Gamma$–$M$ path for selected $N_{k_z}$ values. Near the Fermi energy, the differences between calculations using $N_{k_z} = 1$ and $N_{k_z} = 6$ across constant-$k_z$ planes are essentially negligible. For the bulk compound, our calculations reproduce the metallic character and the main dispersive features reported SARPES measurements, in agreement with previous theoretical studies  ~\cite{Jiang2025NP}. In particular, the relative positions, bandwidths, and Fermi-level crossings of the vanadium-derived bands closely match those extracted from SARPES, confirming that the low-energy electronic structure is faithfully captured without invoking an on-site Hubbard correction. The resulting PAOFLOW bands are indistinguishable from the Quantum ESPRESSO eingenvalues, precisely capturing the gap opening and overall dispersion, providing an identical electronic foundation for the subsequent analysis of altermagnetic magnon excitations.

The band dispersions reflect the characteristic altermagnetic spin splitting: the spin character of the bands is inverted between the $\Gamma$-X and $\Gamma$-Y directions, a band that is predominantly spin-down along $\Gamma$-X becomes predominantly spin-up along $\Gamma$-Y, and vice versa. We have verified that variations of the out-of-plane momentum $k_z$ do not qualitatively modify the in-plane projected band structure, consistent with the quasi-two-dimensional nature of the bulk compound \cite{bf1n-sxdl}.

\begin{figure}[!htbp]
    \includegraphics[width=0.25\textwidth]{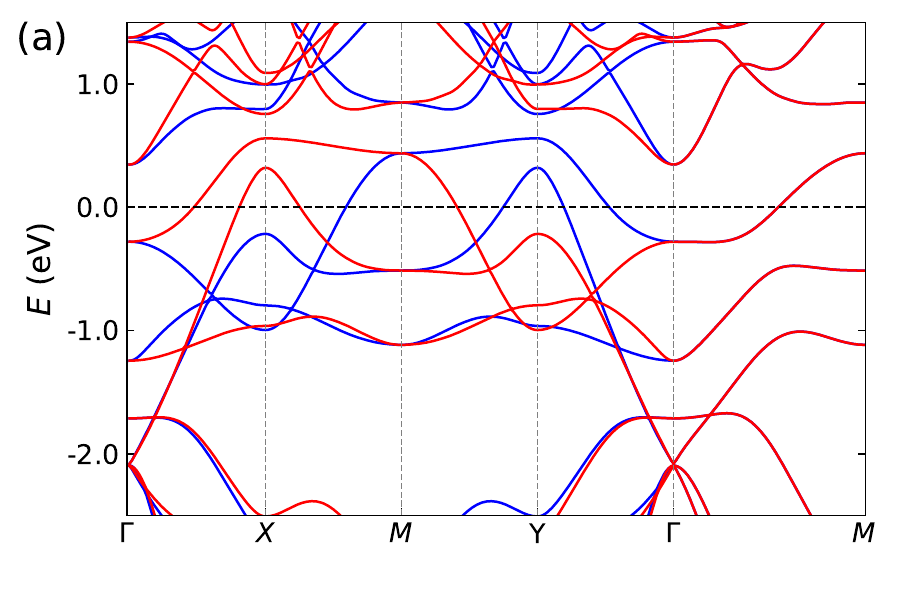}
    \includegraphics[width=0.25\textwidth]{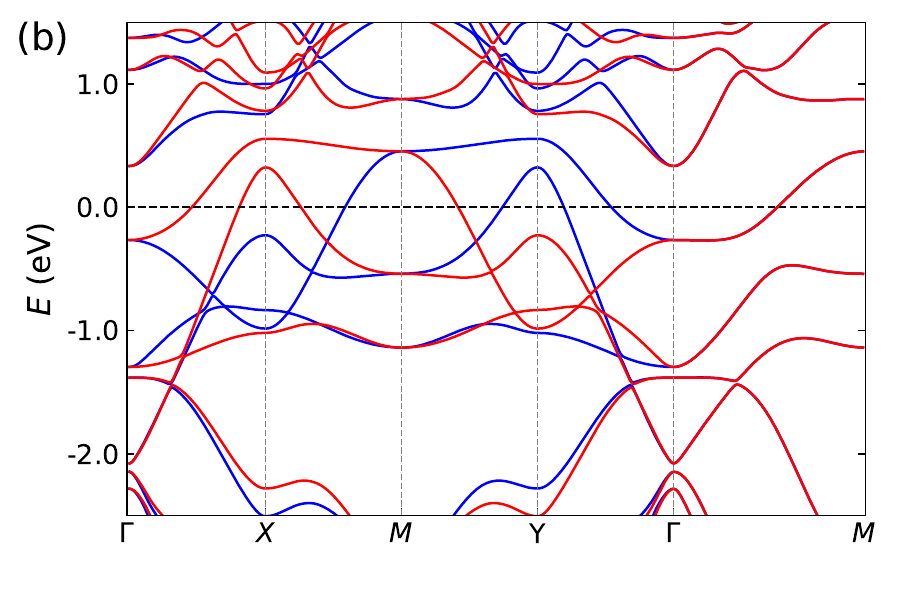}
    \vspace{0.1 cm}
    \includegraphics[width=0.25\textwidth]{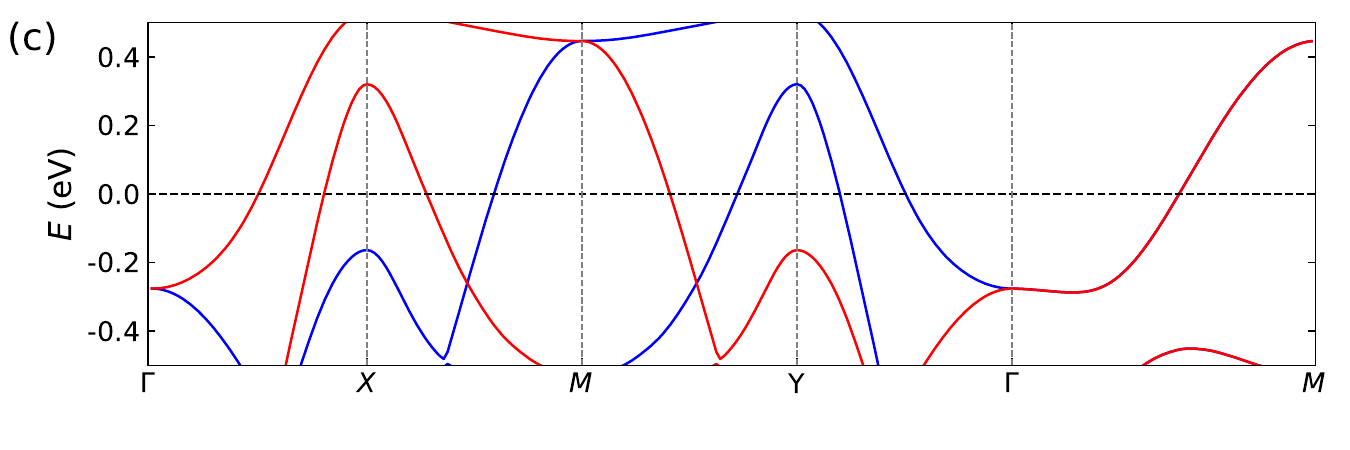}
    \includegraphics[width=0.25\textwidth]{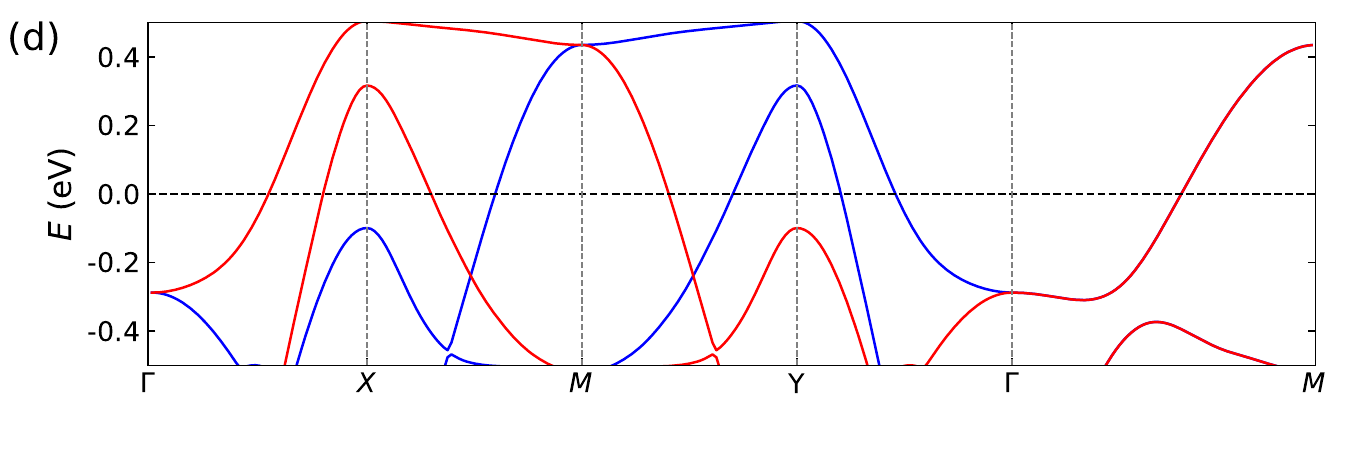} 
        \caption{\label{fig:fig_suppl1} Spin-resolved electronic bands of bulk $\text{KV}_2\text{Se}_2\text{O}$ along the high-symmetry path $\Gamma$–$X$–$M$-$Y$–$\Gamma$-$M$, exhibiting a metallic character. (a) for $Nk_z=1$; for $Nk_z=6$ calculated for plane with (b)$k_z = 0$, zoom close to the Fermi energy for (c)$k_z = 0.25 (2\pi/c)$ and (d) $k_z = 0.5 (2\pi/c)$.}

\end{figure}

Figure~\ref{fig:fig_suppl2} shows the electronic band structure of $\text{KV}_2\text{Se}_2\text{O}$ with and without spin--orbit coupling (SOC). The overall effect of SOC on the electronic structure is relatively weak. The most noticeable modification is the lifting of band crossing points near the $\Gamma$ point as shown in Fig.~\ref{fig:fig_suppl2}. In the vicinity of the Fermi level, which dominates the behavior of the Green's functions and the low-energy spin excitations, the band structure remains largely unaffected. This indicates that SOC does not significantly alter the low-energy electronic states responsible for the magnetic response, but instead introduces subtle symmetry-breaking effects.

\begin{figure}[!htbp]
	\includegraphics[width=0.35\textwidth]{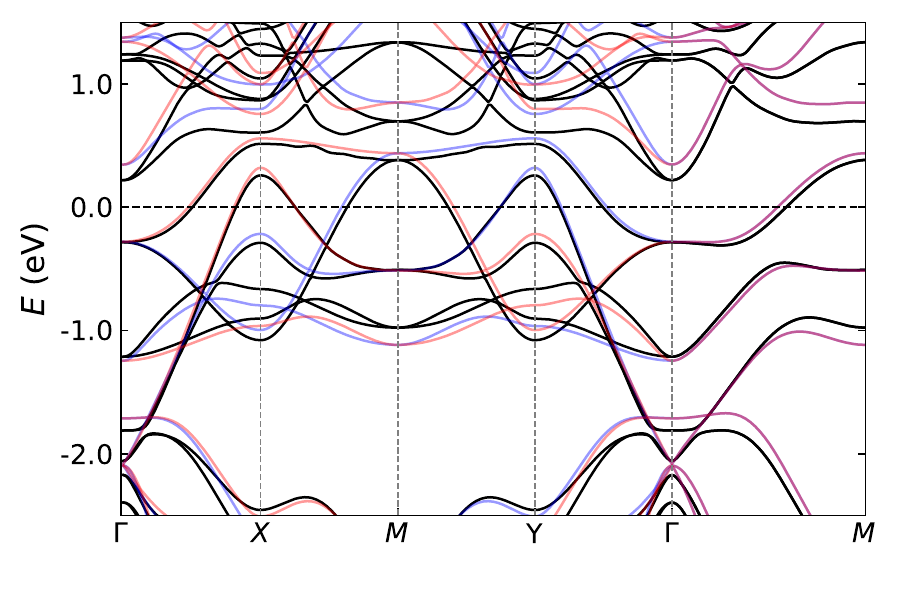}
    \caption{\label{fig:fig_suppl2} (a) Electronic band structure of $\text{KV}_2\text{Se}_2\text{O}$ with (black curves) and without SOC (spin-up is blue shaded curves and spin-down is red shaded curves). (b)Transverse magnon dispersion relation. The two magnon branches with SOC are shown as solid lines (black for the first mode and red for the second mode), while the corresponding dispersions without SOC are represented as shaded curves. }
\end{figure}

\section{Estimation of the transition temperature}

Our estimation of the transition temperature is based on renormalized spin 
wave theory~\cite{TahirKheli1962,Callen1963,Lee1967},
\begin{equation}
    \frac{1}{T_N} = \frac{1}{N}\sum_{\vec{k}}\left( \frac{\xi_1(\vec{k})}{\omega_1(\vec{k})} + \frac{ \xi_2(\vec{k}) }{ \omega_2(\vec{k}) } \right), 
\end{equation}
where $\omega_i(\vec{k})$ are the magnon dispersion relations for the two lowest magnon branches, 
including the anisotropy gap, and $\xi^s_i(\vec{k})$ are the Bogoliubov weights at sublattice $s$ for mode $i$, arising from linear spin wave theory applied to an effective spin Hamiltonian~\cite{Pires2021Book}. By fitting the two 
magnon branches to a spin model on a square lattice with second-neighbor exchange couplings appropriate to the altermagnetic symmetry, (see Fig.~\ref{fig:spinmodel} for details) and evaluating the expression above, we obtain $T_{N}\approx 450\,\mathrm{K}$.

\begin{figure}
    \centering
    \includegraphics[width=0.5\linewidth]{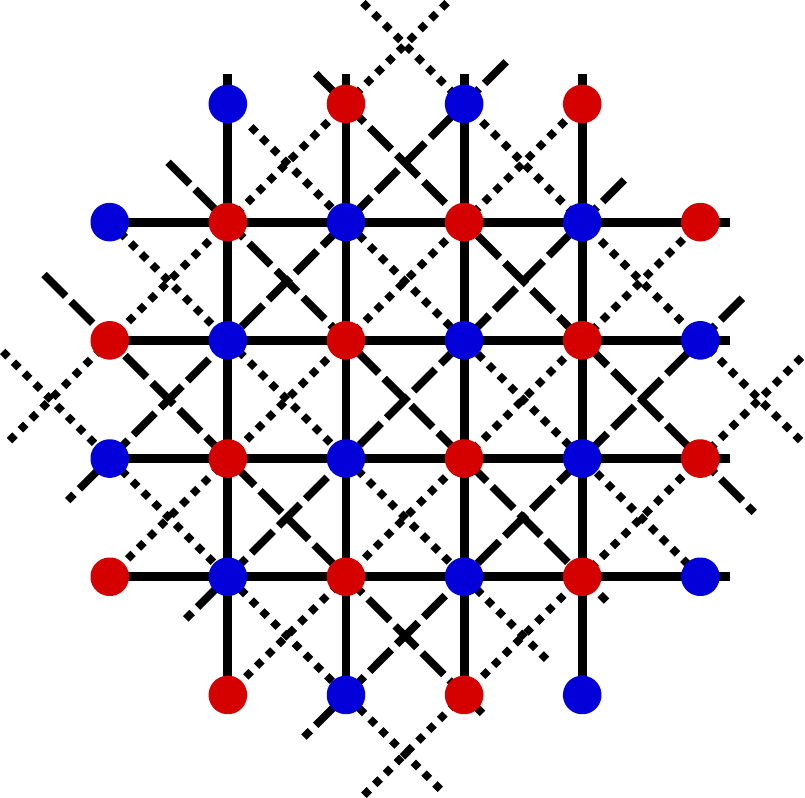}
    \caption{Schematic representation of the effective spin model used in the estimation of the critical temperature $T_N$. The red and blue circles represent the two opposite-spin sublattices, the solid lines represent the nearest-neighbor exchange $J$,
    the long- and short- dashed lines represent the second-neighbor altermagnetic exchanges $J_+$ and $J_-$. Fitting the itinerant magnon dispersion relation calculated through RPA (main text) to such a spin model in the linear spin wave approximation, we obtain $J\approx 161$~meV, $J_+\approx 5.4$~meV and $J_-\approx 19$~meV. The fitting also includes a third-neighbor exchange of $\approx 14$~meV.
    }
    \label{fig:spinmodel}
\end{figure}


\end{document}